# THE RUSSIAN PRACTICE OF APPLYING CLUSTER APPROACH IN REGIONAL DEVELOPMENT

**Victor Grebenik,**
*Professor, Financial University under the Government of the Russian Federation, Moscow, Russia*
**Yuri Tarasenko,**
*Associate Professor, Law Institute of the Russian University of Transport, Moscow, Russia*
**Dmitry Zerkin,**
*Associate Professor, Law Institute of the Russian University of Transport, Moscow, Russia*
**Mattia Masolletti,**
*Lecturer, NUST University, Italy*

**Abstract**

The article considers the practice of applying the cluster approach in Russia as a tool for overcoming economic inequality between Russian regions. The authors of the study analyze the legal framework of cluster policy in Russia, noting that its successful implementation requires a little more time than originally planned. Special attention is paid to the experience of benchmarking.

1. **Introduction**

In order to stimulate the economy and support the regions, the implementation of the cluster model in Russia is one of the tools in the policy of reducing regional inequality. For the successful strategic development of Russia, it is necessary to identify the poles of growth, understand their prospects, and develop development assistance measures that allow spreading the growth impulses of the new economy. The updating of regional strategies will be more effective if the subjects of the federation take into account their role in certain poles of growth and development corridors. This is the basis for the innovative development of regional policy [7].

This goal will be achieved through the implementation of the state regional policy aimed at realizing the development potential of each region, overcoming infrastructure and institutional constraints, creating equal opportunities for citizens and promoting human development, carrying out targeted work on the development of federal relations, as well as reforming the systems of state administration and local self-government.

## 2. Main part

The Russian expert Valitsky defines cluster policy as a system of measures of state support aimed at the development of clusters, which, in turn, ensure the stable and balanced development of regions and the country as a whole, as well as the stimulation of innovation and the realization of competitive advantages, in particular [8].

In the Russian Federation, the practice of applying cluster policy has been applied relatively recently. The clusters that function at this stage of policy development were created on the basis of territorial production complexes that existed in the conditions of a planned economy. Many experts in the fields of economics and regional development believe that cluster policy in Russia is currently not comprehensive, that is, it is not expressed in the form of a specific policy covering a number of sectors of the economy, with a clearly defined strategy, functioning mechanisms and a specific budget.

The cluster approach is one of the priority directions of innovative development of the regions. The authorities have taken significant steps towards the formation of the country's cluster policy, which is confirmed by the adopted legal documents (Fig. 1).

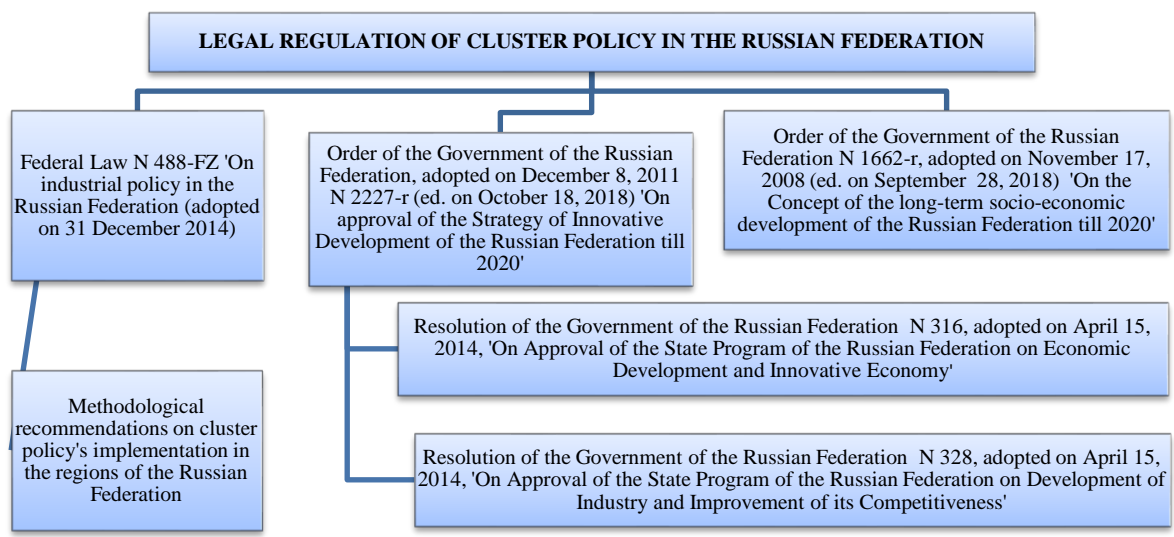

Fig.1. Legal regulation of cluster policy in the Russian Federation

Source: compiled by the authors on the basis of the official website of the Russian Cluster Observatory. URL: https://cluster.hse.ru

In a number of regions, separate programs for the development and support of clusters are being adopted or organizational structures responsible for the development of clusters are being created.

According to the methodological recommendations of the Russian Ministry of Economic Development, 'territorial clusters are an association of enterprises, suppliers of equipment, components, specialized production and service services, research and educational organizations connected by relations of territorial proximity and functional dependence in the field of production and sale of goods and services. At the same time, clusters can be located on the territory of one or several subjects of the Russian Federation' [2].

The basic objectives of the cluster policy are to achieve high rates of economic growth, increase the production functionality of enterprises forming territorial production clusters, and ensure the diversification of the economy due to the growing potential of enterprises.

Practical research in the field of cluster policy is carried out by the Russian Cluster Observatory (hereinafter, the RCO) – regional innovation policy and cluster development, established on the basis of the Institute for Statistical Research and Knowledge Economics of the National Research University 'Higher School of Economics'. Since 2016 the RCO has been contributing to the implementation of the priority project 'Development of Innovative Clusters - Leaders of the World-Class Investment Attractiveness', launched by the Ministry of Economic Development of Russia.

Examples of the formation of new clusters include the creation of the Moscow research and production cluster on the basis of existing facilities of a developed information infrastructure, which will contribute to the growth of the technological potential and investment attractiveness of the capital. The following clusters are also in the process of implementation: in the Omsk region - a cluster for the production and deep processing of milk, in Tatarstan – a pharmaceutical cluster, in Udmurtia - a woodworking cluster. All of these projects will improve the economic situation in the home regions and reduce the level of regional inequality. Cluster policy is quite a new phenomenon in the policy of economic development of the regions of the Russian Federation, it is an important and effective tool that is given considerable attention in the state strategies of regional development at various levels. Thus, cluster policy is an exclusive institutional form of public and private partnership mechanism, as well as a tool for the implementation of regional development [6] strategies and business projects.

The presence of a large number of clusters with a low level of development has demonstrated a poor efficiency of the functioning and development of cluster structures in the country. At the same time, an important factor justifying the current situation is that the majority of clusters (83.3%) had been formed between 2012 and 2016. According to the researchers, to achieve a high level of organizational development, the cluster needs at least 10-15 years [4].

In 2016 the Ministry of Economic Development had launched a priority project, known as 'Development of Innovation clusters'. Its main goal is to create growth points to ensure faster growth rates in the regions where innovative territorial clusters are based on achieving a world-class investment attractiveness, effective support for entrepreneurship, and integration into global value chains.

At the present stage, the country has a large number of clusters that are at the initial level of organizational development, which does not allow us to fully speak about the efficiency of their activities. This measure is quite widespread in developed countries [9]. For instance, it is actively applied in Austria, Belgium, Great Britain, Germany, Italy, France [1], and Sweden.

In Russia, the so-called benchmarking is also used in order to reduce regional inequality.

Benchmarking in a broad sense is the search for the best experience that can be applied in the field of its activity. This tool is more widely used in the commercial sphere, however, in the field of regional management, it is also used taking into account the specifics of each individual region. The use of the best regional experience is a driver for the further development of a particular subject, improving the work of its internal mechanisms.

There are various forms of exchange of best practices: holding forums, compiling collections of best practices, organizing and conducting competitions (contests), direct dialogue between stakeholders. Such a number of forms can be explained by the different nature of a particular practice, its scale, and the scope of its focus.

One way to share best practices is through forums. They are large dialogue platforms, within which there is a possibility of direct dialogue between representatives of government agencies and business. One of the largest forums is the Annual Eastern Economic Forum, which had been established by the Decree of the President of the Russian Federation of No. 250 'On establishment of the Eastern Economic Forum' (adopted on May 19, 2015).

Another way to share best practices is to compile so-called methodological expert collections. As examples of best practices in the field of economics, we can mention the 'Collection of Best Practices for Improving the Investment Climate in the Subjects of the Russian Federation', published and compiled by the Agency for Strategic Initiatives in 2018 (hereinafter, the ASI). In addition to the above mentioned method of identifying best practices, the ASI uses other tools aimed at investigating a narrower area of activity in the region. Another example, a collection of successful practices for implementing a regional investment standard includes the best practices of interaction between regional authorities and investors from different regions, including the Republics of Bashkortostan and Tatarstan, the Leningrad Region, the Tyumen Region, the Khanty-Mansi Autonomous District (geographically included in the Tyumen Region), and others.

In 2014 the Ministry of Economic Development of the Russian Federation held the all-Russian Contest 'The best management solutions of regional authorities for the Development of the Investment Environment'. On the platform of the contest the regions of Russia had presented unique management solutions to create a favorable investment environment [3]. Another example - the Agency for Strategic Initiatives had initiated the All-Russian Competition of Best Practices and Initiatives for the Socio-economic Development of the subjects of the Russian Federation, starting from 2018 and taking place up to the present moment.

As we see, regional benchmarking is a broad phenomenon that encourages regional alignment, since best practices can be found in any industry and field.

**Conclusion**

Thus, the key objectives of Russia's cluster policy are the following: to support the development of new industries (information technologies [5], biotechnologies, new materials, etc.) by creating an innovative environment around them; to increase the competitiveness of small and medium-sized enterprises; to update old-industrial agglomerations and regions. The region, in which the pilot innovation territorial cluster operates, undoubtedly acquires a number of significant advantages for further socio-economic development in comparison with other territories due to a number of indicators that significantly exceed the average values for their home regions.